# Non-magnetic Pulsar Emission Mechanism

Nikolay Kamardin

## Abstract

It is generally accepted that the pulsar magnetic field converts most of its rotational energy losses into radiation. In this paper, we propose an alternative emission mechanism, in which neither the pulsar's rotational energy nor its magnetic field is involved. The emission mechanism proposed here is based on the hypothesis that the pulsar matter is stable only when moving with respect to the ambient medium at a velocity exceeding some threshold value. A decrease in velocity below this threshold value leads to the decay of matter with the emission of electromagnetic radiation. It is shown that decay regions on the pulsar surface in which the velocities of pulsar particles drops to arbitrarily small values are formed under simple driving condition. It is also shown that for the majority of pulsars having measured transverse velocities, such a condition is quite possible. Thus, the pulsar radiation carries away not the pulsar rotational energy, but its mass, while the magnitude of the rotational energy does not play any role. At the end of the paper, we consider the reason for the possible short-period precession of the pulsar.

## 1. Introduction

The foundations of the modern understanding of the nature of pulsars emerged shortly after the discovery in 1967 of the first space object emitting radio signals with an extraordinarily stable pulsation period. The discoverers of the new object suggested [1] that the source of such a signal is a white dwarf or neutron star, the radiation of which is modulated by the radial pulsations of the star. Although this hypothesis of radial pulsations has been addressed in other works (for example, [10, 11]), it has not received further confirmation.

According to T.Gold [2], the source of a pulsating radio signal can only be a rotating neutron star (NS), which has regions of different radiation intensity. The structure of the observed pulses represents directional beams rotating like a lighthouse beacon, and depends only on the distribution of emission regions around the circumference of the star. It is the rotation of the massive body that maintains the high stability of the pulsation period. The enormous rotational energy of the pulsar is transformed gradually into radiation; therefore the rotation speed is observed to decrease with time, albeit very slowly.

A substantial contribution to the initial theory of pulsars was made by F. Pacini [3], who introduced the oblique rotator model to the theory. In this model, a NS with a powerful dipolar magnetic field rotates around an axis that does not coincide with the magnetic axis. Such a configuration successfully explains the periodic motion in space of the radiation beam generated by the magnetic field. Only in this case (the presence of a nonzero angle between the magnetic axis and the axis of rotation) can a rotating magnetic dipole in a vacuum emit energy [4]. Due to these two properties, the oblique rotator model has become the dominant theoretical pulsar model since then [5].

The existence of celestial bodies, which can be assumed to be the sources of the pulsating signals, was predicted well before pulsars were discovered. As early as 1932, L. Landau [6] suggested the existence of superdense stars in which the nuclei of atoms are densely packed. Two years later, after the discovery of the neutron, W.Baade and F.Zwicky [7] called such objects neutron stars, and were the first to indicate the conditions for their appearance. They proposed that a superdense body of a very small radius, consisting mainly of neutrons, could result from a supernova core collapse and, as a result, the former core of the star becomes an independent celestial object. Somewhat later, it was substantiated [8, 9] that the neutron core is formed in fairly massive stars in the last stage of their life cycle. Thus, the discovery of pulsars indirectly confirmed the existence of the NS's and, at the same time, raised many questions concerning their properties and the properties of the interstellar medium. One of the most difficult and still unresolved issues is the pulsar emission mechanism in a broad frequency range from IR to hard X-rays.

In many works on pulsars, from early papers [12, 13] to recent ones (see, for example, [14, 15, 16, 17]), a key role in the pulsar emission mechanism is attributed to the superstrong magnetic field, which is believed to be possessed by every NS. However, the processes occurring in neutron star magnetospheres described in the works above do not exclude another possible explanation of the pulsar emission activity.

In this paper, we consider an alternative pulsar emission mechanism that is not associated with NS magnetic fields. The proposed mechanism is based on the hypothesis that the NS matter is stable only when moving at high velocity with respect to the ambient medium. If the matter particles' velocity decreases to a certain threshold value $v_{th}$, they decay into radiation.

This paper shows that, with certain ratios of the spatial and angular velocities of the pulsar, an area is formed in its body where a radiation beam can be formed.

## 2. Kinematic properties of pulsar

Pulsars possess a unique combination of short rotation periods (from milliseconds to tens of seconds) and high space velocities (50 to 1000 km/s). They are thought to acquire rapid rotation rates at the moment of their birth in supernova explosions [18] when the cores of massive aging stars collapse. In this case, their initial rotation speed increases by orders of magnitude due to conservation of angular momentum.

In contrast, the issue of the reason for the high space velocities of pulsars remains open. Since NSs' velocities are much higher than those of ordinary stars (10 to 20 km/s), it is assumed that there is some mechanism that accelerates these compact objects during or after their formation. There are two main competing mechanisms that could accelerate NSs during their birth: anisotropic ejection of the stellar debris ("hydrodynamic kicks" [19, 20]) or asymmetric neutrino emission ("neutrino-induced kicks" [21, 22]). These mechanisms work during the first few seconds after a supernova explosion. Another mechanism involves the existence of topological vector currents [23], which slowly but steadily accelerate a star throughout its life.

The pulsar's spin period is determined through the precise measurement of pulse arrival times and is known with high accuracy for almost all detected pulsars. At the same time, transverse velocities are determined for a much smaller number of pulsars, since this requires additional measurements of their proper motions. Therefore, in the ATNF Pulsar Catalogue[1] [24], only 14.4% of pulsars have transverse velocity values at the time of writing. Unfortunately, large errors in the determination of the pulsar distances (about 25% [25]) significantly reduce the accuracy of the velocity measurements. Besides, the transverse velocity of a pulsar can be considered as a projection of its space velocity on the plane of the sky perpendicular to the direction of the pulsar and can only serve as a lower estimate of the magnitude of pulsar space velocity.

[1] http://www.atnf.csiro.au/research/pulsar/psrcat v1.64

## 3. Velocity of points of a pulsar. Instantaneous axis of rotation

Assuming the pulsar to be a solid homogeneous spherical body, consider its motion. It rotates about its axis and moves through the interstellar medium. This motion is described by the vector $\overline{v}$ of the velocity of the center of mass O, and the vector $\overline{\omega}$ of the angular velocity (aligned along the axis of rotation). Denote the angle between these vectors by $\alpha$ $(0 \le \alpha \le 180°$, and can be a periodic function of time in the case of precession of the axis of rotation). In the case of this motion, the velocities of the pulsar's points change and are different for different points (if these points do not lie on one straight line parallel to the axis of rotation). We decompose the vector $\overline{v}$ into a sum of two orthogonal components $\overline{v}_{\parallel}$ and $\overline{v}_{\perp}$, where the vector $\overline{v}_{\parallel}$ is parallel and the vector $\overline{v}_{\perp}$ is normal to the axis of rotation (see Fig. 1). Thus we separate the motion of the pulsar into planar ($\overline{v}_{\perp}, \overline{\omega}$) and axial ($\overline{v}_{\parallel}$) motion.

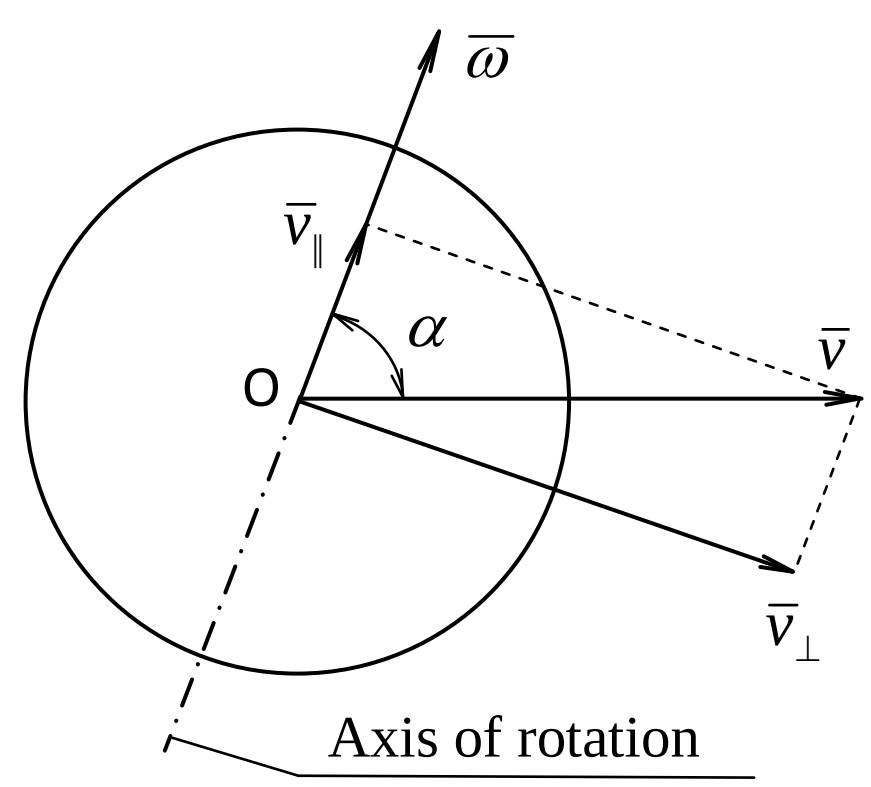

Fig.1. Pulsar motion.

Consider separately the pulsar's planar motion described by the vectors $\overline{v}_{\perp}$ and $\overline{\omega}$. According to Euler's theorem of classical mechanics, this complex motion can be considered as the sequence of instantaneous rotations of the body with the same angular velocity ω about an instantaneous axis of rotation, which is parallel to the body axis of rotation. The distance between the body axis of rotation and the instantaneous axis of rotation is

$$e = v\sin(\alpha)/\omega \qquad (1)$$

and their common plane is perpendicular to the vector $\overline{v}_{\perp}$.

If we consider the rotation around the instantaneous axis of rotation, the motion of an arbitrary point M on the body can be treated as a composition of its rotation around the instantaneous axis and the motion parallel to this axis along with the body. The absolute velocity of this point is expressed as follows:

$$v_M = \sqrt{\omega^2 l_M^2 + v^2\cos^2(\alpha)}\ , \qquad (2)$$

where $l_M$ is the distance from point M to the instantaneous axis of rotation.

Consider a pulsar of radius R. If the quantities ω, $v$ and α are such that the condition

$$e < R \qquad (3)$$

is satisfied, then the instantaneous axis of rotation crosses the body of the star. In this case, the points of the body located on the instantaneous axis of rotation ($l_M = 0$) have an arbitrarily small absolute velocity if the angle $\alpha$ close to 90º. In this regard, we verify the fulfillment of condition (3) for pulsars from the ATNF catalog. For this purpose, we take the absolute velocity of the pulsar equal to its transverse velocity, $R = 10$ km and $\alpha = 90°$. We rewrite condition (3) in the notation adopted in the ATNF pulsar catalog:

$$V_{trans} P0/(2\pi) < 10 \ , \qquad (4)$$

where $V_{trans}$ is the transverse velocity of the pulsar (km/s) and $P0$ is the period of rotation (s). At the time of writing, we have found that out of 412 pulsars having measured transverse velocities, 256 satisfy condition (4).

Since the instantaneous axis of rotation does not coincide with the axis of rotation, the distance $l_M$ of an arbitrary point M to the instantaneous axis changes cyclically during each revolution of the pulsar, if this point does not lie on the axis of rotation (see Fig. 2). In this case, the magnitude of the absolute velocity of point M also changes cyclically, according to formula (2). The minimum velocity is achieved at $l_M = | e - r_M |$, where $r_M$ is the distance from point M to the axis of rotation of the pulsar.

Let us suppose that the instantaneous axis of rotation crosses the body of the star, and consider this question: Which of its points has the minimum absolute velocity that is less than a given value of $v_{th}$? It turns out that these are the points that approach the instantaneous axis of rotation at a distance less than some value of $s$ at each revolution of the pulsar. In other words, these are the points that in their movement intersect a cylinder of radius $s$, whose axis is the instantaneous axis of rotation.

We find the value of $s$ from the condition that the absolute velocity of the points on the surface of the cylinder is equal to $v_{th}$:

$$v_{th} = \sqrt{\omega^2 s^2 + v^2 \cos^2(\alpha)}\ , \text{ whence}$$

$$s = \frac{\sqrt{v_{th}^2 - v^2 \cos^2(\alpha)}}{\omega} \qquad (5)$$

Therefore the answer to the above question is: points whose distance $r$ to the axis of rotation of the pulsar satisfy the condition:

$$|e - r| < s \qquad (6)$$

have a minimum value of absolute velocity less than a given value of $v_{th}$. Since the cylinder of radius $s$ consists of points whose absolute velocity is lower than $v_{th}$, we will call it the Low-Velocities Cylinder (LVC). The time spent by a point in the LVC depends on how close to the instantaneous axis of rotation its trajectory passes. The point M whose trajectory intersects the instantaneous axis of rotation has the maximum time $\tau_{\max}$ spent inside the LVC (see Fig. 2).

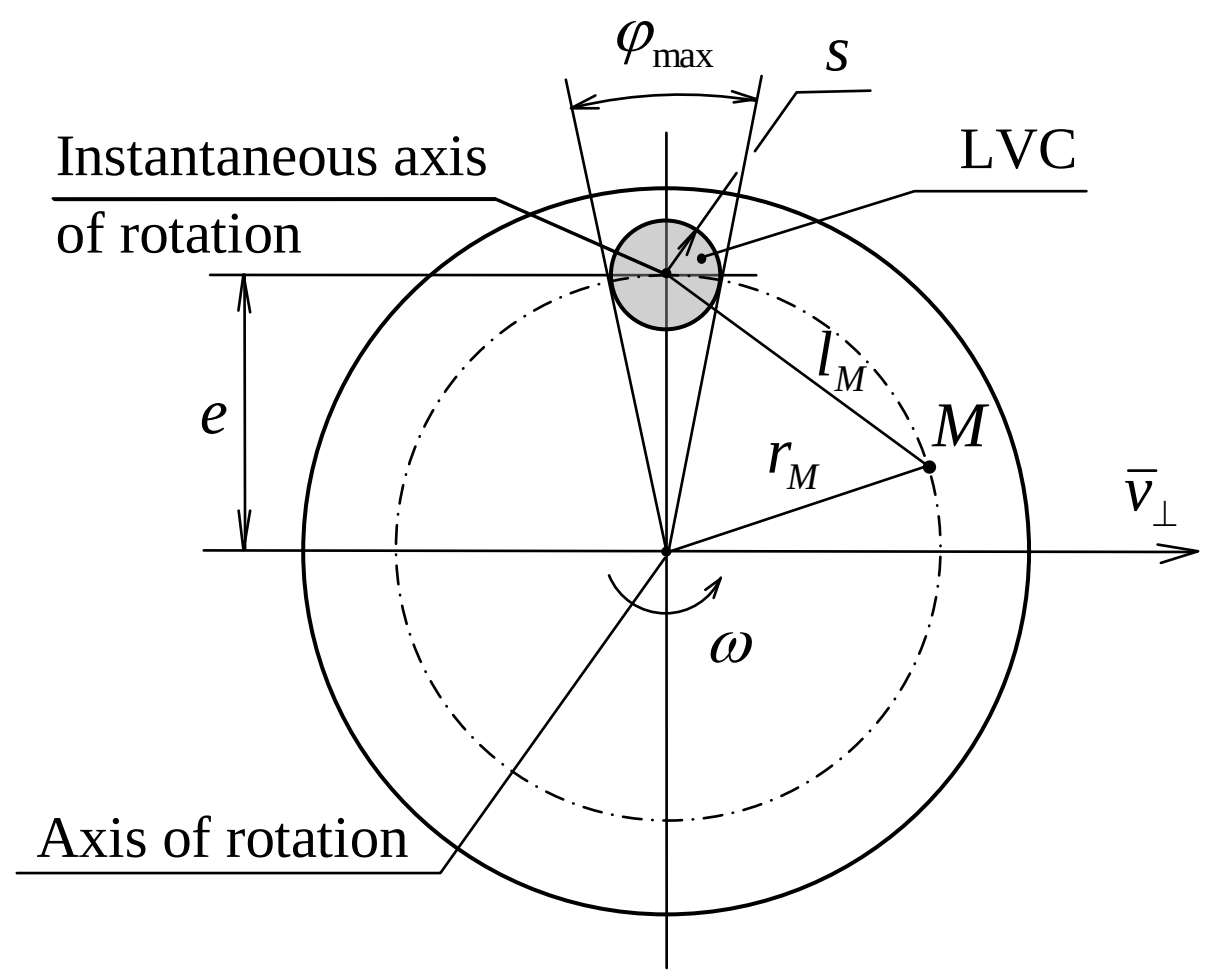

Fig. 2. View of the pulsar in the direction of the axis of rotation. $M$ is a point whose trajectory intersects the instantaneous axis. While the point $M$ is moving inside the LVC, the pulsar rotates through the angle $\varphi_{\max}$.

We find the expression for $\tau_{\max}$. While the point M is moving inside the LVC, the pulsar rotates through the angle

$$\varphi_{\max} = 2\arcsin\left(\frac{s}{e}\right) = 2\arcsin\left(\frac{\sqrt{v_{th}^2 - v^2\cos^2(\alpha)}}{v\sin(\alpha)}\right), \quad (7)$$

then the required time is

$$\tau_{\max} = \frac{\varphi_{\max}}{\omega} = \frac{2}{\omega}\arcsin\left(\frac{\sqrt{v_{th}^2 - v^2\cos^2(\alpha)}}{v\sin(\alpha)}\right) \quad (8)$$

## 4. Pulsar emission mechanism

As already mentioned, the proposed model is based on the hypothesis that the pulsar matter is stable only when moving with respect to the ambient medium at a velocity exceeding a certain threshold value $v_{th}$. A decrease in velocity below this value leads to the decay of pulsar matter with the emission of electromagnetic radiation.

In the previous section, we have shown that with a certain ratio of angular to linear velocity (if the angle $\alpha$ between their vectors is close to $90°$), in the pulsar a region LVC forms, in which the absolute velocity of the points is less than $v_{th}$. The application of the above hypothesis to this case allows us to propose a new emission mechanism of the pulsar not considering the effect of magnetic fields.

### 4.1. Particle velocity

Without considering the structure of NS matter, we combine in the concept of “particle” any smallest part of matter, which can move independently of others. For example, in a Coulomb crystal, the particle is an ion located in the lattice site and experiencing thermal oscillations. In a neutron liquid, the particles are neutrons and other elementary particles (if any) that are in a stable state.

Choose a rotating coordinate system so that the origin is at the NS center of mass, and one of the axes is directed along the axis of rotation. The particle velocity in the body-fixed frame we will call the *internal velocity*. This velocity has a key feature - if the particle does not participate in the vortex motion, there are time instants at which this velocity vanishes. For such a particle it is therefore possible to indicate time periods during which the modulus of its internal velocity is smaller than any predetermined value.

The absolute velocity of any particle has the form

$$\bar{v}_{abs} = \bar{v}_{\text{int}} + \bar{v}_M , \qquad (9)$$

where $\bar{v}_{\text{int}}$ is the internal velocity and $\bar{v}_M$ the transport velocity (velocity of the point M which is associated with the NS and where the particle is at the moment).

### 4.2. Particle decay

As noted above, when the instantaneous axis of rotation crosses the pulsar, an LVC with a radius $s$ corresponding to the value of $v_{th}$ is formed around this axis. NS matter in this region slows down so that the transport velocity $v_M$ of any particle becomes lower than $v_{th}$. At the same time, among these particles there are those for which the internal velocity modulus $v_{\text{int}}$ is so small that the condition

$$\left| \bar{v}_{\text{int}} + \bar{v}_M \right| < v_{th} \qquad (10)$$

is fulfilled for them. If this condition persists for a sufficient time, then the particle begins to decay. We assume that if the particle decay has begun, it cannot be interrupted or stopped when the particle leaves the zone where the decay has begun, and must go to completion under any conditions. Therefore, if the particle decay time $\tau_{dec}$ exceeds the time spent by the particle in the LVC, it completes its decay outside the LVC zone.

The portion of the pulsar surface on which the particles are in a state of decay, we will call the *decay region*. Thus, the decay region is a part of the pulsar surface cut out by the surfaces of coaxial cylinders of radius $e-s$ and $e+s$ and two cylindrical surfaces of radius s (see Fig. 3). The angular extent of this area is equal $\omega\tau_{dec}$. On the pulsar surface, there are two decay regions which are symmetric with respect to the equatorial plane. Although the body rotates, the decay regions do not change position with respect to the plane passing through the pulsar axis of rotation and its instantaneous axis of rotation.

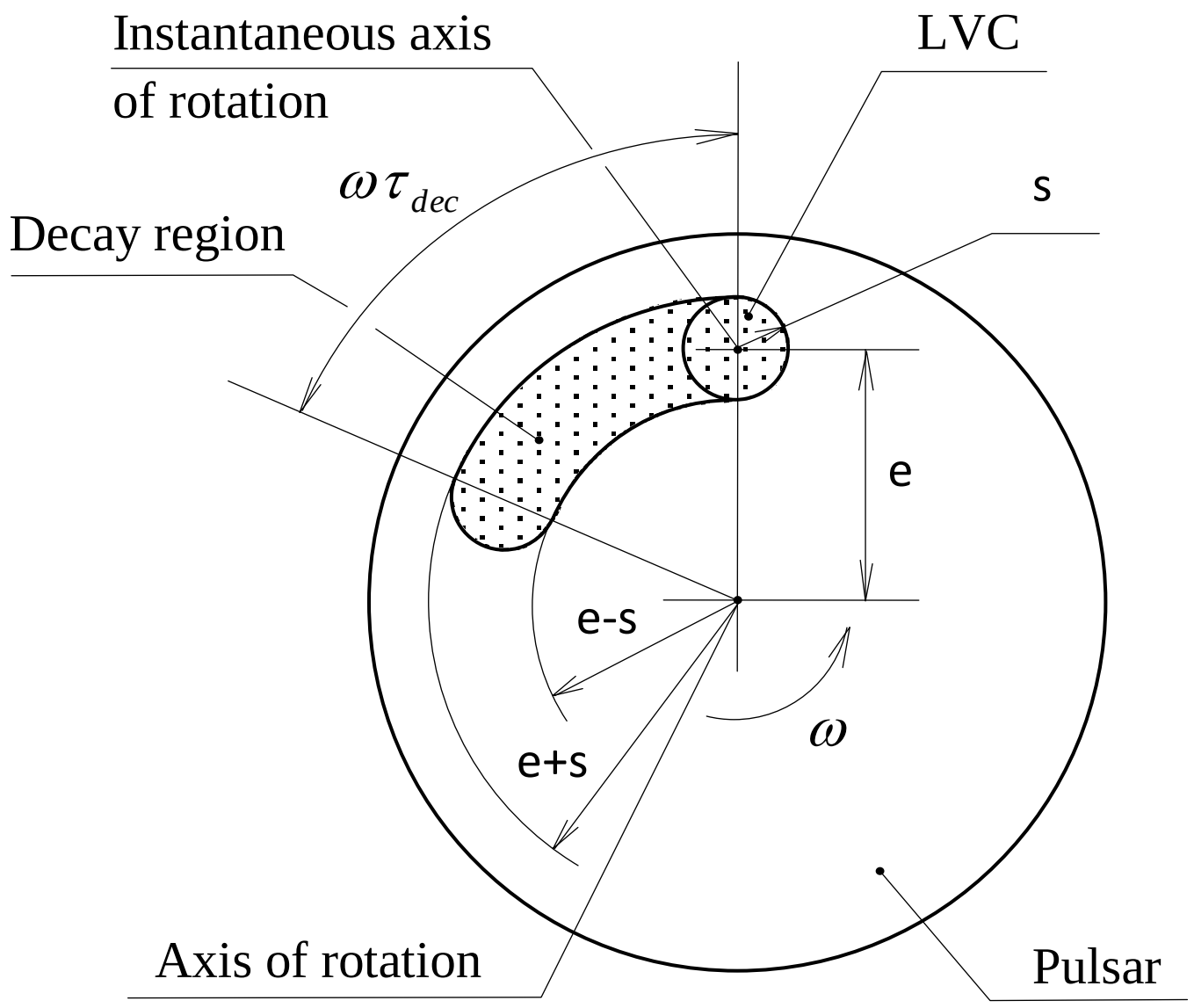

Fig. 3. The decay region on the pulsar surface. View in the direction of the axis of rotation of the body. $\tau_{dec}$ is the particle decay time.

## 4.3. Formation of pulsar radiation

The decay regions on the pulsar surface are those hot spots from which the radiation emanates. The shapes and locations of the decay regions described in Section 4.2 are consistent with the results of recent studies described in [26, 27] of two independent research groups. In these works, the authors investigate a variety of emission patterns on the surface of the millisecond pulsar PSR J0030+0451 using a Bayesian inference approach to analyze its energy-dependent thermal X-ray waveform, which was observed using the Neutron Star Interior Composition Explorer (NICER). It is shown that, contrary to expectations, the spots pattern is clearly not that of a centered dipole and the hot spots have the shapes of oval, or an azimuthally extended crescent, that supports the idea of the non-magnetic nature of the pulsar emission. Otherwise, the star must have a magnetic field of an unusually complex configuration, which is not consistent with the spherically symmetric models of the body.

The electromagnetic radiation generated in the decay regions leaves the surface of the pulsar in the form of two oppositely directed outflow jets. It should be noted that with the described decay regions locations on the pulsar surface, the radiation fluxes propagate symmetrically with respect to the equatorial plane and the directions of these fluxes make an angle between themselves less than $180°$.

Here we present an illustrative calculation of the energy release associated with the decay of matter in the decay regions using the example of an isolated millisecond pulsar PSR J2235 + 1506. This pulsar has a pulse period of $P0 = 0.0598$ s and transverse velocity $V_{trans} = 131.62\ km/s$ . We assume that the space velocity of the pulsar is equal to its transverse velocity $v_p = V_{trans}$, and the angle between the angular and linear velocity vectors $\alpha = 90°$. We take the threshold value of velocity $v_{th} = 1\ km/s$. The distance between the axis of rotation of the pulsar and the instantaneous axis of rotation:

$$e = \frac{v_p\, P0}{2\pi} = \frac{131.62*10^3*0.0598}{2\pi} = 1252.7\, m$$

LVC radius:
$$s = \frac{v_{th}\, P0}{2\pi} = \frac{1*10^3*0.0598}{2\pi} = 9.5\ m$$

We assume that the outer crust of a NS consists of a body-centered cubic lattice of iron $^{56}Fe$ [28] with a density of $\rho = 8.02*10^9\ kg/m^3$, with the depth of the decay regions $h = 1*10^{-6}\ m$. The volume of NS matter passing through the decay regions for one rotation of the star:

$$V_1 = 2\pi e\ 2s\, 2h = 2\pi*1252.7*2*9.5*2*10^{-6} = 0.30\ \ m^3$$

Suppose that only one particle out of 1000 decays in the volume $V_1$, $k = 1/1000$. Then the amount of energy released in the decay regions for 1 s is:

$$W = \frac{V_1\, \rho\, k\, c^2}{P0} = \frac{0.30*8.02*10^9*1/1000*(3*10^8)^2}{0.0598} = 3.62*10^{24}\, J/s = 3.62*10^{31} erg/s$$

Next, we estimate the time required for the complete decay of the pulsar. We take the mass of the pulsar to be
$$M_n = 1.4\, M_{Sun} = 1.4*2*10^{30} = 2.8*10^{30}\ kg$$

Amount of matter decaying for one rotation of the star:

$$M_1 = V_1\, \rho\, k = 0.30*8.02*10^9*1/1000 = 2.4*10^6\ kg$$

For simplicity, we assume that the decay rate is constant over a pulsar's lifetime. Then the total pulsar's lifetime is approximately

$$T_n = \frac{M_n}{M_1} P0 = \frac{2.8*10^{30}}{2.4*10^6} 0.0598 = 7*10^{22}\ s = 2.2*10^{15}\ yr$$

## 5. Pulsar axis precession

It is believed that pulsar emissions are produced within its magnetosphere, which co-rotates with the pulsar like a rigid body, and the emission regions remain stationary relative to the surface of the pulsar and rotate with it. Such a model clearly explains the periodic motion in space of the pulsar emission beam when the magnetic axis of the star is oblique with respect to the rotational axis. In this case, the main pulsation periods of the signal correspond to the pulsar rotation periods.

In the non-magnetic mechanism of pulsar emission, the emission regions do not rotate with the pulsar, but remain stationary relative to the plane passing through the pulsar spin axis and its instantaneous axis of rotation (see 4.2). If so, then the periodic motion in space of the emission beam can be explained by the precession of the pulsar spin axis, and the periods of the signal correspond to the precession periods.

However, if we consider a pulsar as a homogeneous spherical body (or a body composed of multiple concentric spherical layers), which rotates around the center of mass without external forces, it should not have a precession. Such a rotation corresponds to the Euler case of the rigid body motion in which the spin axis maintains a constant direction in space.

Nevertheless, there are several observations of periodic changes in pulse shape and/or arrival times of pulses that seem to indicate the presence of precession. In these cases, the precession period is estimated to be in the range from several hours [29, 30] to hundreds of days [31, 32, 33, 34]. It is believed that the precession arises from the aspherical deformation of the star under the influence of strong magnetic fields.

In this paper, we propose a new, completely different explanation for pulsar precession. In the emission mechanism presented here, the star has an unusual property - it permanently changes its configuration during rotation. That is, the body itself rotates, while the decay regions in which the mass loss occurs remain stationary. The rotational stability of such a body requires careful study, but for now we can evaluate its stability by comparing the main moments of inertia.

In the Appendix, we consider a pulsar model that shows that the value of the principal moment of inertia $I_\varsigma$ of the body relative to the rotation axis remains constant, while the values of other two, $I_\xi$ and $I_\eta$ , change cyclically twice per rotation with small but equal amplitudes and are in antiphase with each other.

Thus, there are body positions in which one of the following conditions holds:

$$\begin{aligned} I_\eta < I_\varsigma < I_\xi \\ I_\xi < I_\varsigma < I_\eta \end{aligned} \qquad (11)$$

This means that the value of the moment of inertia $I_\varsigma$ becomes intermediate between $I_\xi$ and $I_\eta$ four times per rotation (see Fig. A2), but it is known from classical mechanics [35] that in the Euler case, the permanent rotation around such an axis is unstable. We suggest that this instability is the main mechanism responsible for the short-period precession of the pulsar spin axis.

102/2 apt. 28 Leninsky Prospect street,
Voronezh, Russia.
E-mail address: kavict@yandex.ru

## Appendix: Determination of the principal moments of inertia of a rotating uniform sphere, from the surface of which material is ejected

Consider a rotating uniform sphere of radius $R$ and mass $M$, on the surface of which there are two identical compact emitting regions from which there is a continuous ejection of material. The average size $d$ of the emitting regions is assumed to be small ($d << R$). A straight line passing through the midpoints of these regions is at a distance $e$ from the center of the sphere. The body rotates about a central axis parallel to this line.

The emitting regions have a peculiar feature – they do not maintain a fixed position on the surface of the rotating body, but remain stationary, sliding on its surface.

We choose a body-fixed Cartesian coordinate system $\xi$, $\eta$, $\varsigma$ whose origin is at the center of the body, and the axis $\varsigma$ is directed along the axis of rotation. We denote by $\varphi$ the angle of body rotation - as shown in Fig. A1. For simplicity, we assume that the emitting regions do not emit material continuously but just at the beginning of each revolution - at $\varphi = 0$, each emitting region instantly loses a mass m of matter ($m << M$). Given the smallness of m, we can neglect the displacements of the center of mass of the body relative to the center of the sphere.

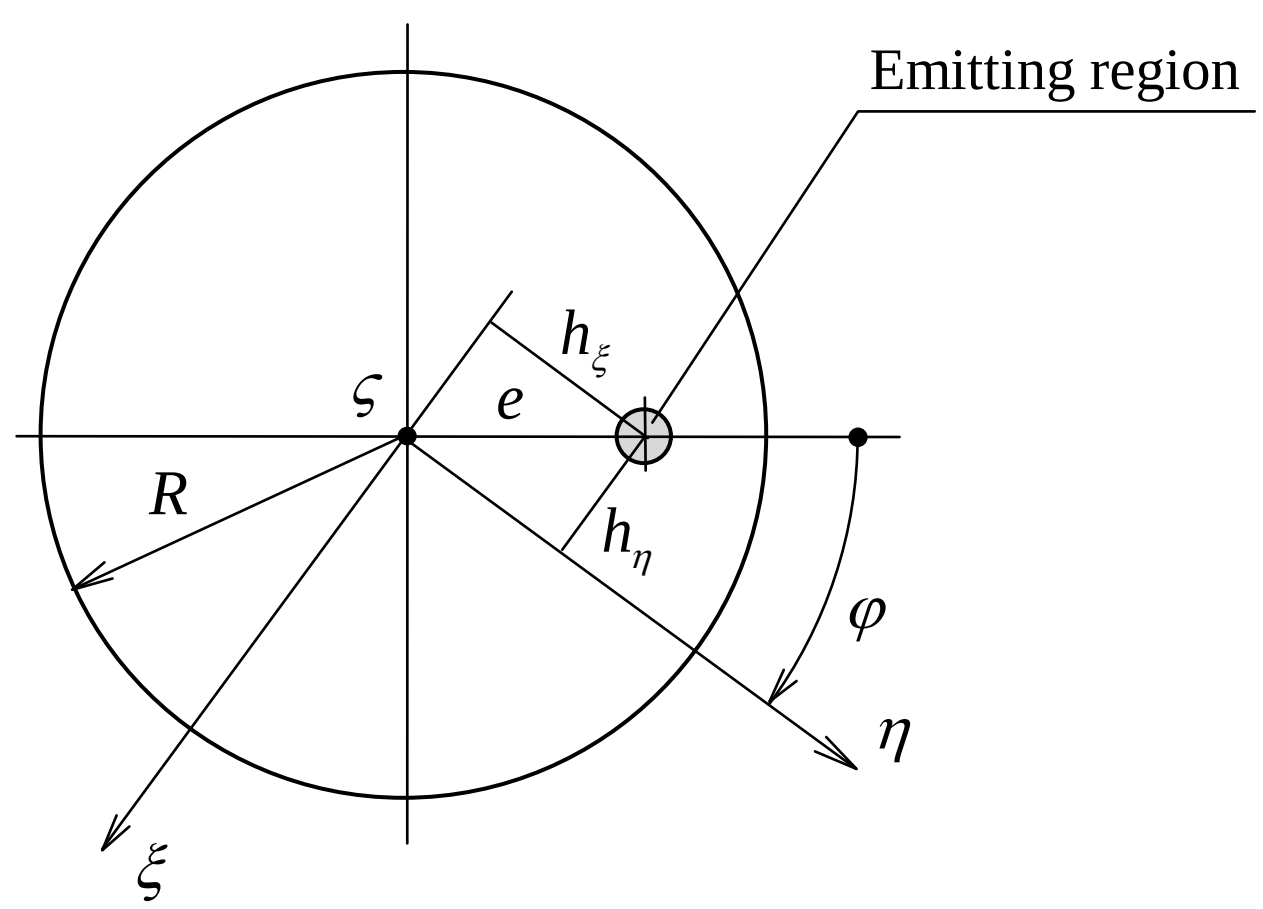

Fig. A1. Rotating uniform sphere. View in the direction of the axis of rotation. $\xi$, $\eta$, $\varsigma$ - body-fixed Cartesian coordinate system.

The distance between the midpoints of the emitting regions:

$$l = 2\sqrt{R^2 - e^2} \qquad (1)$$

The moment of inertia of the body, relative to the rotation axis:

$$I_\varsigma = \frac{2}{5}MR^2 - 2me^2 \qquad (2)$$

The moments of inertia of the considered body, relative to the $\xi$ and $\eta$ axes:

$$\begin{aligned} I_\xi &= \frac{2}{5}MR^2 - 2m\left(\left(\frac{l}{2}\right)^2 + h_\xi^2\right) \\ I_\eta &= \frac{2}{5}MR^2 - 2m\left(\left(\frac{l}{2}\right)^2 + h_\eta^2\right) \end{aligned} \qquad (3),$$

where $h_\xi = e\cos(\varphi)$, $h_\eta = e\sin(\varphi)$ (4) are the distances from the line passing through the midpoints of the emitting regions to the coordinate axes (see Fig. A1). Substituting (1) and (4) into (3), we obtain:

$$\begin{aligned} I_\xi &= \frac{2}{5}MR^2 - 2m\left(R^2 - e^2 + e^2\cos^2(\varphi)\right) \\ I_\eta &= \frac{2}{5}MR^2 - 2m\left(R^2 - e^2 + e^2\sin^2(\varphi)\right) \end{aligned} \qquad (5)$$

We consider qualitatively how the ratios of the main moments of inertia change as the body rotates. In expressions (2) and (5), we omit identical terms $\frac{2}{5}MR^2$, adopt for simplicity the value $m = 1/2$, and we derive the following expressions for the variable components of the main moments of inertia:

$$\begin{aligned} \bar{I}_\xi &= -\left(R^2 - e^2\sin^2(\varphi)\right) \\ \bar{I}_\eta &= -\left(R^2 - e^2\cos^2(\varphi)\right) \\ \bar{I}_\varsigma &= -e^2 \end{aligned} \qquad (6)$$

These curves for $e/R = 0.72,\ 0.82,\ 0.95$ are shown in Fig. A2. It can be seen that there are the angles of body rotation (indicated by segments with arrows) on which one of the conditions holds:

$$\begin{aligned} I_\eta < I_\varsigma < I_\xi \\ I_\xi < I_\varsigma < I_\eta \end{aligned}$$

That is, in these positions, the value of the moment of inertia of the body relative to the rotation axis falls between the values of the two other main moments of inertia. According to the rotational dynamics of a rigid body, in these positions the rotation is unstable.

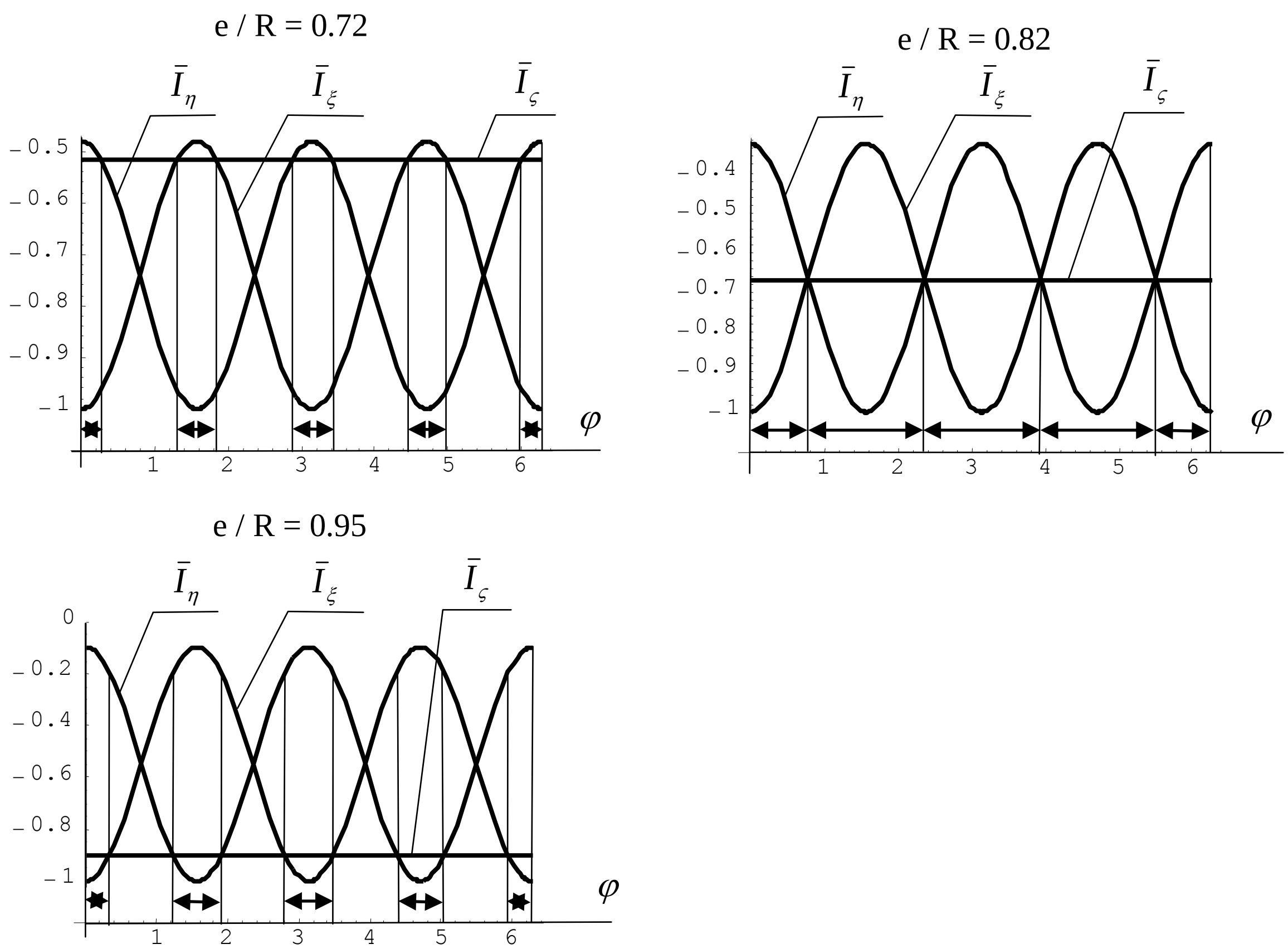

Fig. A2. Variable components of the main moments of inertia as a function of the angle of body rotation at various values $e/R$. The segments with arrows denote the sections of the angle of rotation on which rotation is unstable.